\documentclass[aps,prb,preprint,showpacs,preprintnumbers,amsmath,amssymb]{revtex4}
\usepackage{epsf}

\begin{document}

\newcommand{\ve}[1]{\boldsymbol{#1}}

\title{Proton Wires in an Electric Field: the Impact of
Grotthuss Mechanism on Charge Translocation}
\author{Natalie Pavlenko\footnote[1]{Present address: Institut f\"{u}r Theoretische Physik,
Physik-Department der TU M\"{u}nchen, James-Franck-Strasse,
D-85747 Garching b.~M\"{u}nchen, Germany;\\
E-mail: natalie@physik.tu-muenchen.de}}
\address{Institute for Condensed Matter Physics, 1 Svientsitsky
St., UA-79011, Lviv, Ukraine}

\begin{abstract}
We present the results of the modeling of proton translocation in
finite H-bonded chains in the framework of two-stage proton
transport model. We explore the influence of reorientation motion
of protons, as well as the effect of electric field and proton
correlations on system dynamics. An increase of the reorientation
energy results in the transition of proton charge from the
surrounding to the inner water molecules in the chain. Proton
migration along the chain in an external electric field has a
step-like character, proceeding with the occurrence of electric
field threshold-type effects and drastic redistribution of proton
charge. Electric field applied to correlated chains induces first
a formation of ordered dipole structures for lower field strength,
and than, with a further field strength increase, a stabilization
of states with Bjerrum D-defects. We analyze the main factors
responsible for the formation/annihilation of Bjerrum defects
showing the strong influence of the complex interplay between
reorientation energy, electric field and temperature in the
dynamics of proton wire.

\end{abstract}



\maketitle

\section{INTRODUCTION}

Translocation of protons over long distances has a key importance
for biological and chemical systems. It is believed that proton
migration along the chains of water molecules formed between the
interior of proteins and the solvent, establishes electrochemical
potential gradients playing an important functional role
\cite{gennis,lanyi}. Experimental evidence indicates that the dominant
mechanism responsible for proton transport in transmembrane
proteins (for instance, in bacteriorhodopsin of {\it Halobacterium
halobium} \cite{stoeckenius,nagle} and in gramicidin A channels
\cite{levitt,nagle}) is the diffusion of H$^+$ ions which is
faster than the hydrodynamic flow of hydronium species
(H$_3$O)$^+$. Especially at low hydrogen concentrations in
channels, proton conduction is determined by a two-stage
Grotthuss-type mechanism \cite{cukierman,onsager} shown
schematically in Fig.~1(a). The first stage involves the intrabond
proton tunnelling along the hydrogen bridge which is connected
with the formation and transfer of ionic positive (H$_3$O$^+$) and
negative (OH$^-$) charged defects. To sustain a flux of H$^+$ in
such proton wire, the inter-molecular proton transfer due to the
reorientations of molecular group with proton is assumed.
Reorientation motion leads to the breaking of the hydrogen bond
(so-called orientational Bjerrum L-defect) and location of proton
between another pair of molecular groups \cite{bjerrum}.
Consequently, the reorientation step in the presence of the second
proton may induce high-energy configuration with both of the
protons shared by two adjacent oxygen ions (Bjerrum D-defect).

Unlike the translocation of monovalent ions like Cs$^+$, Na$^+$ or K$^+$
via gramicidin requiring the net diffusion of the whole water column in the channel,
the existence of the Grotthuss-type selective migration of H$^+$ through the H-bonded chain
is supported by the absence of streaming potentials during H$^+$ permeation \cite{akeson,levitt}.
In contrast to the bulk water, the reorientation motion in one-dimensional water wire involving a migration of Bjerrum defects with the period of reorientations about
$10^{-10}$~s, is much slower than the proton intra-bond hopping ($\sim
10^{-12}$~s)\cite{nagle1,chernyshev}. This is also closely
related to the fact that the mobility of Bjerrum defects ($\sim
10^{-4}$ cm$^3$V$^{-1}$s$^{-1}$) is much lower than that of the
ionic defects ($\sim 10^{-2}$ cm$^3$V$^{-1}$s$^{-1}$)
\cite{eisenberg}. Moreover, as appears from the results of molecular dynamics simulations \cite{pomes3},
the translocation of the ionic defects in proton wires is almost activationless process, whereas 
the orientation defects involving an activation energy of about 5~kcal/mol in the chain containing eight water molecules, constitute a limiting step for the proton migration. 
Besides the orientation defects, the recent experiments indicate that another possible
rate-limiting step for the proton migration in gramicidin channels can be at the membrane-channel/solution interface\cite{chernyshev,godoy}.

As was pointed in \cite{cukierman}, the experimental analysis of the
proton flow in bioenergetic proteins and the mechanism of proton
translocation is very difficult because of its intrinsically transient
nature. To shed more light on the microscopic nature of the proton
transport and to analyze the influence of quantum effects and
interaction with proton surrounding, theoretical modeling remains
to be essentially important. Recently, much attention has been
focused on the theoretical studies of the dynamics of ionic defects
using soliton models \cite{zol1,zol2} and 
molecular dynamics simulations \cite{tuck1,tuck2,schumaker,schumaker2}. Proton transfer
in water was shown to be strongly coupled with the dynamics of local
environment, and the density of ionic defects was found to increase
exponentially with the increasing temperature\cite{zol1}. However, since
the concentration of slow Bjerrum defects in water solutions is
much higher ($c_B=2 \cdot 10^{-7}$ at $-10 ^{\circ}$~C) than that
of fast ionic ($c_I=3 \cdot 10^{-12}$)\cite{eisenberg}, the
investigations of the reorientation step of proton migration are
necessary for the better understanding of the proton transport process.

The goal of the present work is to study proton wire containing a
finite number of water molecules by the use of quantum statistical
mechanics methods which are extremely effective in the description
of the collective nature of the proton transfer and in the quantum treatment
of the light H nuclei \cite{davydov2}. To describe correctly the proton
transport process, we employ here two-stage proton transport model
\cite{jps} incorporating quantum effects such as proton tunneling
and zero-point vibration energy. In earlier papers
\cite{pavlenko1,pavlenko2,pavlenko3} we applied the two-stage model to
analyze the effect of coupling between protons and molecular group
vibrations on proton conductivity in infinite H-bonded chains and
proton-conducting planes. In particular, it was shown that the
proton-lattice vibration interactions induce structural
instabilities and charge ordering in system \cite{pavlenko1},
whereas the Grotthuss-type transport mechanism manifests itself in
nontrivial temperature- and frequency dependences of the proton
conductivity \cite{pavlenko2,pavlenko3,pavlenko4}.

In this work we analyze the influence of proton-proton
correlations, comparing two different protonated chains containing
one and two excess protons respectively.  We find that the
reaction of protonation of water chain is extremely sensitive to
the reorientation energy barrier of proton motion and the barrier 
for the chain protonation. We show that the
increase of the reorientation energy results in the drastic decrease
of the proton charge density at the boundary between the chain and
surrounding with consequent localization of protons near the inner
water molecules. As appears from our modeling, the application of
external electric field induces the step-like threshold-type
effects with the ordering of proton charge and stabilization of
Bjerrum D-defects in the wire. We analyze the temperature
dependence of proton polarization and D-defect concentration, and
examine the role of the interplay between different factors (such as
orientation energy, external field and temperature) in the
dynamics of Bjerrum defects.

\section{\label{model} MODEL SPECIFICATION AND METHOD OF CALCULATION}

To model a proton wire, we consider a linear chain containing $N$
hydrogen bonds and $l=1,2,\ldots, N, N+1$ molecular groups. The
outer left ($l=1$) and right ($l=N+1$) molecular complexes mimic
the surrounding of the proton wire and differ from the inner
($l=2,\ldots, N$) water molecules. The transport
of an excess proton through the wire proceeds via the following two steps:\\
(i) proton can be transferred within a H-bond (process shown by
short arrows in Fig.~1(a)) which is modelled by a simple double-well
potential, with the corresponding energy barrier $\Omega_T$ for the proton transfer
between the two minima:
\begin{equation}
H_T=\Omega_T \sum_{l=1}^N (c_{la}^+ c_{lb}+ c_{lb}^+ c_{la}),
\label{h_t}
\end{equation}
where $c_{l\nu}^+$($c_{l\nu}$) are the operators of the proton
creation(annihilation) in the position ($l$, $\nu$) (the index
$\nu=\{a ,b\}$
denotes the left/right position for the proton within the H-bond);\\
(ii) a water molecule together with covalently bonded hydrogen ion
can be rotated, and this process causes the breaking of the H-bond
and location of H$^+$ between two another nearest water molecules
of the wire (process depicted by long arrows in Fig.~1(a)):
\begin{equation}
H_R=\Omega_R \sum_{l=1}^N (c_{l+1,a}^+ c_{lb}+ c_{lb}^+
c_{l+1,a}), \label{h_R}
\end{equation}
where $\Omega_R$ is the effective energy barrier for the proton hopping
between the states $|l+1,a \rangle$ and $|l,b \rangle$
(reorientation of the $l$-th molecular group together with
proton). As is shown in \cite{pomes2} by the computation of the proton
mean-force potential, this transition between donor-acceptor and
acceptor-donor states reverses the chain dipole moment and
requires a substantial energy barrier about $5.5$~kcal/mol for the
whole chain containing up to eight water molecules.

Besides the transport process, we incorporate the following two
types of interactions between protons in the chain:\\
(iii) different short-range configurations of the protons near an
inner water molecule as well as an outer molecular group can
appear due to the different nature of bonding (shorter covalent or
longer H-bond). The energies of possible configurations (shown in
Fig.~1(b)) are described by the following terms:
\begin{eqnarray}
&&H_1=\tilde{\varepsilon}_1 (1-n_{1a})+\tilde{w}_1 n_{1a}, \quad
H_{N+1}=\tilde{\varepsilon}_{N+1} (1-n_{Nb})+\tilde{w}_{N+1} n_{Nb}, \label{h_conf}\\
&&H_l=w'(1-n_{l+1,a})(1-n_{lb})+w n_{l+1,a}n_{lb}+ \varepsilon
(1-n_{lb})n_{l+1,a}+\varepsilon n_{lb} (1-n_{l+1,a}).\nonumber
\end{eqnarray}
The parts $H_1$ and $H_{N+1}$ describe the energies of the
boundary proton configurations near the left and the right
surrounding molecular groups (we assume for simplicity
$\tilde{\varepsilon}_1=\tilde{\varepsilon}_{N+1}=\tilde{\varepsilon}$
and $\tilde{w}_1=\tilde{w}_{N+1}=\tilde{w}$ in the boundary
configurations shown in the upper scheme of Fig.~1(b)). The terms
$H_l$ ($l=2,\ldots,N$) contain the configuration energies for the
water molecules in the interior of the wire (the lower part in Fig.~1(b)).
Here the proton occupancy operators $n_{l\nu}=c_{l\nu}^+ c_{l\nu}=\{0,1\}$;\\
(iv) a strong repulsion between two nearest protons shared by two
neighboring oxygens (so called Bjerrum D-defect) with a repulsion
energy $U$ is represented by the term:
\begin{equation}
H_C=U\sum_{l=1}^N n_{la}n_{lb}. \label{h_hubb}
\end{equation}
In our following analysis we use the value of $U\approx
10$~kcal/mol corresponding to the energy of relaxed $D$-defect
estimated in \cite{campbell} on the basis of quantum chemical
calculations.

To model a field exerted by the surrounding, we apply an external
electric field of a strength $E$ to the proton wire, which is
described by the following term
\begin{equation}
-e_p E\sum_{l=1\atop{\nu=\{a,b\}}}^N R_{l\nu} n_{l\nu},
\label{h_e}
\end{equation}
where $R_{l\nu}$ is the coordinate of the proton position ($l$,
$\nu$) with respect to the center of the chain, and $e_p$ denotes
the proton charge.

In order to analyze the dynamics of the proton wire embedded in
the surrounding under the influence of the field, as well as the
effect of rotational motions of covalent groups with proton, we
will focus our attention on the polarization of the proton wire
defined here as
\begin{equation}
P=e_p \sum_{l=1\atop{\nu=\{a,b\}}}^N R_{l\nu} \langle n_{l\nu}
\rangle, \label{pp}
\end{equation}
where $\langle \ldots \rangle$ denotes the statistical average with
respect to the system energy (\ref{h_t}-\ref{h_e}). The average
probabilities $\langle n_{l\nu} \rangle$ of occupation of the
position ($l$, $\nu$) by proton describe the distribution of the
proton charge in the wire, and thus is another very important
characteristics to track the proton migration.

To calculate {\it exactly} the above-mentioned statistical
averages, we need to know the quantum energy levels determined by
the energy (\ref{h_t}-\ref{h_e}). This can be done by a mapping of
the proton states ($l$,$\nu$) on the multi-site basis $|i
\rangle=|n_{1a}, n_{1b}, \ldots, n_{Nb} \rangle$. Then, using the
projection operators $X^{ii'}=|i \rangle \langle i'|$ acting on
the new basis $|i \rangle$ we rewrite the system energy
(\ref{h_t}-\ref{h_e}) in a convenient form (see Appendix):
\begin{equation}
{H}={H^0}\oplus{H^1}\oplus \cdots \oplus {H^{2N}}. \label{ham}
\end{equation}
Each term ${H^{n_p}}$ in (\ref{ham}) corresponds exactly to $n_p$
protons in the chain  ($n_p=\sum_{l,\nu} \langle n_{l\nu}
\rangle=0,1,\ldots, 2N$). This means in fact that the mapping on
the states $|i \rangle$ allows to decompose the terms
(\ref{h_t}-\ref{h_e}) and analyze the cases of different number of
protons in the wire separately.

The energy barrier for the protonation of the chain is described
by the parameter $\Delta$ which appears in (\ref{ham}) after the
decomposition procedure (see Appendix). As follows from the
definition (\ref{delta}), $\Delta$ is the difference between the
energies of the proton attraction to the boundary and to the inner
water molecules. As the PMF-studies of the H-bonded chain dynamics
\cite{pomes2} show that the inner H-bonds are stronger (shorter
O-O separation distances) than the outer H-bonds, it is reasonable
to consider below the case $\Delta<0$ (we take
$|\Delta|=0.85$~kcal/mol in our numerical calculations), when the
proton is attracted to the surrounding and needs to overcome the
boundary energy barrier $|\Delta|$ to protonate the water chain.

The parameter $J=w+w'-2\varepsilon$ (see Fig.~1(b)) is related to
the effective short-range interactions between the protons near the
water molecule. It describes, in fact, the energy of the formation
of the pair of ionic defects: I$_+$=H$_3$O$^+$ ($w-\varepsilon$)
and I$_-$=OH$^-$ ($\varepsilon-w'$) from two water molecules at
the dissociation reaction (2H$_2$O $\rightarrow$
H$_3$O$^+$+OH$^-$). Since the value of $J$ is about $22$~kcal/mol
\cite{eisenberg,rao} and is more than twice as much as $U$, we
exclude in our following analysis an appearance of the pair of
ionic defects in the system.

Due to the two types of motions we have two different
contributions to the proton dipole moment: the orientational part
$\mu_{r}=e_{p}R_{r}$ and the transfer part $\mu_{ab}=e_{p}R_{ab}$
where $R_{ab}=R_{lb}-R_{la}$ denotes the distance H-H between the two
nearest proton positions of the double-well H-bond. In our calculations, we
use the values $\mu_{ab}=3.5$~D and $\mu_{ab}=4.5$~D corresponding
to the moderately strong H-bond with an O-O distance
$R_{ab}+2R_{r}=2.6\AA$ and the covalent O-H bond of a length
$R_r=0.94\AA$.

\section{\label{1-prot}CASE STUDY: ONE EXCESS PROTON IN CHAIN}

As the starting point, in this section we mimic the situation when
one proton is moved towards the water chain embedded into the
solvent. To examine the behavior of the protonated chain with $N$
H-bonds and $n=1$ excess proton, we consider ${H^{1}}$ part of the
energy given by (\ref{ham_decomp}). Since the zero-point vibration
energy for protonated chains is larger than the potential energy
barrier for the proton transfer between two shared oxygens
\cite{pomes}, the quantum tunneling is not required for the intra-bond
H$^+$ transfer. Thus, in our modelling we set $\Omega_T \approx
0$. With this assumption, the energy levels of ${H^{1}}$ can be
found {\it exactly}:
\begin{eqnarray}
\lambda_{1,2}&=&\Delta
\pm\left((N-1)\mu_r+\frac{N}{2}\mu_{ab}\right)E,\nonumber \\
\lambda_{3,\ldots,2N}&=& \left\{
\begin{array}{ll}
\pm i(\frac{{\textstyle\mu_{ab}}}{{\textstyle 2}}+\mu_r)E\pm p
\quad (i=1,3,\ldots,N-2),
\quad $for odd$ \quad N \\
\pm i(\frac{{\textstyle\mu_{ab}}}{{\textstyle 2}}+\mu_r)E\pm p
\quad (i=0,2,\ldots,N-2), \quad $for even$ \quad N
\end{array} \right.
\end{eqnarray}
\noindent where $p=\sqrt{\mu_r^2E^2+\Omega_R^2}$.

To analyze the role of $\Omega_R$ we consider first the case
without external field ($E=0$). Depending on the value of
$\Omega_R$, two different regimes may be stabilized in the system.
In the first {\it small-$\Omega_R$ regime}, the two lowest energy
levels $\lambda_{1,2}$ correspond to the superposition of the two
boundary states
\begin{eqnarray}
|10 \ldots 00 \rangle \quad {\rm and} && \quad |00 \ldots 01 \rangle
\label{outer}
\end{eqnarray}
with the proton located in the surrounding near the left or the
right outer molecular groups. In the second {\it large-$\Omega_R$
regime} the proton is shared between the inner water molecules of
the chain and the ground state of the system corresponds to the
superposition of the states
\begin{eqnarray} |010 \ldots 00 \rangle, |0010 \ldots 00 \rangle
\quad ,\ldots, && |00 \ldots 010 \rangle \label{inner}
\end{eqnarray}
with the energies $\lambda_{4},\lambda_{6},\ldots,\lambda_{2N}$.
The "critical" value $\Omega_R^*=-\Delta$ separating these two
regimes, reflects the transition of the proton from the
surrounding to the states where the proton is shared by the chain
water molecules, which corresponds to the protonation chemical
reaction. Fig.~2 shows the variation of the average occupancies of
proton sites with $\Omega_R$ for the chains containing $N=2$ and $N=3$
H-bonds. For low temperatures (see the case $T=30K$) the boundary
proton occupancies $\langle n_{1a}\rangle =\langle n_{Nb}\rangle$
drop drastically to zero at $\Omega_R=\Omega_R^*$, whereas the
occupation numbers of the central positions $\langle n_{1b}
\rangle=\cdots=\langle n_{Na}\rangle$ increase up to the value
$\frac{1}{2(N-1)}$, reflecting the redistribution of collectivized
proton between the inner sites in the wire. It should be noted
here that $\Omega_R^*$ reflects the change of the ground state of
the system and is determined as the solution of the equation
$\lambda_{1,2}=\lambda_{4,6,\cdots,2N}$ at $E=0$ which does not
depend on temperature. However, as all statistical averages, the
average proton occupancies (for example, of the states $|1,b
\rangle=\frac{1}{\sqrt{2}}(\tilde{|1,b \rangle}+\tilde{|2,a
\rangle})$ and $|2,a \rangle=\frac{1}{\sqrt{2}}(-\tilde{|1,b
\rangle}+\tilde{|2,a \rangle})$ where $\tilde{|1,b \rangle}$ and
$\tilde{|2,a \rangle}$ are the diagonalized states corresponding
to $\lambda_4$ and $\lambda_3$ respectively), are temperature
dependent. Thus the value of $\Omega_R^*(T)\approx-\Delta+kT \ln
2$ where $\langle n_{1a}\rangle =\langle n_{2b}\rangle$, for $T\ne
0$ is not equal to $\Omega_R^*$ (see Fig.~2, case $T=300K$). This
difference shows that the inner proton states (\ref{inner}) are
stabilized already at lower $\Omega_R=\Omega_R^*<\Omega_R^*(T)$,
although the occupancies of the inner positions at
$\Omega_R=\Omega_R^*$ are still slightly lower then of the outer
due to the temperature-induced fluctuations. As $T \rightarrow 0$,
the fluctuations decrease and $\Omega_R^*(T) \rightarrow
\Omega_R^*$.

The effect of the proton localization inside the chain is
supported by the results reported in \cite{pomes,pomes2} showing
that in H-bonded finite chains, the excess charge is best solvated
by the central H-bonds. However, as results from the presented above analysis, the effect of
proton localization is drastically influenced by the competition between two different
tendencies: (i) for small $\Omega_R$, the proton is located near the surrounding/wire interface
due to the nonzero protonation barrier $\Delta$; (ii) to
overcome the barrier between the surrounding and the wire, the
reorientation energy should be sufficiently large ($\Omega_R >
-\Delta$) in order to stabilize the inner proton configurations.
These conclusions show that in general, these two different factors (interface barrier and
orientations) can be rate-limiting for the charge translocation and proton conductivity
of the wire. As was shown in \cite{phillips}, the effective reorientation barrier can be
influenced by temperature factor or applied voltage (for example, the reorientation rate of the wire 
decreases exponentially with $T$ decrease).
Thus, one can also expect that the increase of the temperature can result in the lower
orientation barrier $\Omega_R$, delocalization of proton and consequently in higher values for the proton conductivity through the channel. However, more detailed theoretical analysis 
is needed to understand better the role of the interface in the behavior of the conductivity. 

We turn now to an analysis of the proton translocation directed by
the external field (the case $E \neq 0$). Fig.~3 shows the
field-dependences of $P$ and $\langle n_{l\nu} \rangle$ for the
chain containing $N=2$ H-bonds. We note that the behavior in the
first small- and in the second large-$\Omega_R$ regime differs
drastically. In the small-$\Omega_R$ regime the polarization
increases smoothly with $E$ approaching finally its maximal
saturation value $P_{max}=\mu_r+\mu_{ab}$ (Fig.~3(a), inset). In
contrast to this, in the large-$\Omega_R$ regime the field
dependence is rather nontrivial: first, for $E<E_{thresh}$, $P$
changes very slowly, and than, at $E \sim E_{thresh}$ a strong
increase of $P$ to $P=P_{max}$ is observed in Fig.~3(a). This
rapid step-like change of the proton polarization reflects the
threshold-type effect where the threshold electric field value at
low $T$ is given by
\begin{eqnarray}
E_{thresh}=\frac{\Delta\cdot \mu_H+\sqrt{\Delta^2\mu_r^2+\Omega_r^2
(\mu_H^2-\mu_r^2)}}{\mu_H^2-\mu_r^2}, \quad (\mu_H=\mu_r+\mu_{ab})
\label{thresh}
\end{eqnarray}
and does not depend on the chain size $N$ ($E_{thresh}\approx
0.15\cdot 10^{-7}$~V/cm for the chains with $N=2$ as can be
observed in Fig.~3(a)). As we see in Fig.~3(b), the proton charge
translocation under the influence of the field in this case proceeds
not smoothly, but has a step-like character. As results from
(\ref{thresh}), the threshold value $E_{thresh}$ (which is needed
to overcome a barrier for pumping between the inner localized
states (\ref{inner}) and the boundary state $|00 \ldots 01
\rangle$ in the direction of field) increases for larger
$\Omega_R$ (Fig.~3(a), the cases with $\Omega_R=1.5$~kcal/mol and
$\Omega_R=2.15$~kcal/mol). This implies that the conductivity of
protonated chains can drop with an increasing $\Omega_R$ which can
occur in system for example due to the temperature-induced
fluctuations of $\Omega_R$. However, as was shown in \cite{phillips}, the reorientation rate
increases at increased voltage, which corresponds in our case to the smaller
values of $\Omega_R$ for $E \neq 0$. Thus, we can expect that the external field-induced lowering
of the orientation barrier for the proton translocation results in the increase of
the proton conductivity in the wire.

The drastic change of $P$ at $E=E_{thresh}$ leads to a strong
qualitative difference in the temperature shapes of the polarization
profiles shown in Fig.~4(a) for the large-$\Omega_R$ regime. For
$E<E_{thresh}$ and $\Omega_R>\Omega_R^*$, the excess proton is
located in the central sites, and the increase of $T$ results in a
disorder-induced transfer of the proton from the inner positions
to the chain boundary giving the increase of $P$ at $300$~K as
compared to $50$~K (Fig.~4(a), cases $E=0.02\cdot 10^7$~V/cm and
$E=0.1\cdot 10^7$~V/cm). As the proton is located in the outer
state $|00\ldots 01 \rangle$ for $E>E_{thresh}$ (corresponding to
$P=1$ for $T \rightarrow 0$), the dominant effect of $T$ in this
case is the the disorder-induced proton redistribution between all
sites in the chain leading to the lowering of total polarization
(Fig.~4(a), cases $E=0.2\cdot 10^7$~V/cm and $E=0.3\cdot
10^7$~V/cm). The profiles of the polarization for $\Omega_R$ below
$\Omega_R^*$ shown in Fig.~4(b) appear to be very similar to the
high-field profiles in Fig.~4(a). Since for $\Omega_R<\Omega_R^*$
the proton is located in the outer states (\ref{outer}) near the
chain boundary already at low $E$, the increase of $T$ suppresses
the polarization due to the increasing proton disorder.

\section{ROLE OF PROTON-PROTON CORRELATIONS}

In order to examine the influence of proton correlations, we
consider next the translocation of two excess protons in the wire which
is described by the part ${H}^2$ of the total energy
(\ref{ham_decomp}). Since the presence of two protons in wire may
lead to the formation of Bjerrum D-defect, the energy ${H}^2$ for
the chain with $N=2$ H-bonds given by the expression
(\ref{ham22}), contains the terms with the energy $U$ of the
repulsion between two nearest-neighboring protons.

Analogously to the 1-proton wire, we analyze first the behavior of
the chain without the electric field. For $N=2$ and $\Omega_T
\approx 0$ the energy levels found from (\ref{ham22}) have the
following form:
\begin{eqnarray}
&& \lambda_{1,2}=\frac{1}{2}(U\pm q_{-})+\mu_H E, \nonumber  \label{U-levels}\\
&& \lambda_{3}=\Delta, \quad \lambda_{4}=-\Delta+J, \\
&& \lambda_{5,6}=\frac{1}{2}(U\pm q_{+})-\mu_H E, \qquad
(q_{\pm}=\sqrt{(U \mp 2 \mu_r E)^2+4\Omega_R^2}) \nonumber
\end{eqnarray}
and correspond to the following states of the wire:
\begin{eqnarray} \label{U-states}
&&|\tilde{1}\rangle=\frac{p_{-}}{\sqrt{\Omega_R^2+p_{-}^2}}
|6\rangle +
\frac{\Omega_R}{\sqrt{\Omega_R^2+p_{-}^2}} |7\rangle , \nonumber \\
&&|\tilde{2}\rangle=\frac{\Omega_R}{\sqrt{\Omega_R^2+p_{-}^2}}
|6\rangle -
\frac{p_{-}}{\sqrt{\Omega_R^2+p_{-}^2}} |7\rangle , \nonumber \\
&&|\tilde{3}\rangle=|8\rangle, \quad |\tilde{4}\rangle=|9\rangle, \\
&&|\tilde{5}\rangle=\frac{\Omega_R}{\sqrt{\Omega_R^2+p_{+}^2}}
|10\rangle +
\frac{p_{+}}{\sqrt{\Omega_R^2+p_{+}^2}} |11\rangle , \nonumber \\
&&|\tilde{6}\rangle=-\frac{p_{+}}{\sqrt{\Omega_R^2+p_{+}^2}}
|10\rangle + \frac{\Omega_R}{\sqrt{\Omega_R^2+p_{+}^2}} |11\rangle
\nonumber
\end{eqnarray}
with $p_{\pm}=\frac{1}{2}(U\pm q_{\pm})\mp \mu_rE$.

To study the influence of $\Omega_R$, we analyze (\ref{U-levels})
and (\ref{U-states}) for $E=0$ assuming $J \gg U$ and neglecting
in this way by the formation of ionic defects described by the
configuration $|9\rangle=|0110\rangle$. Similarly to the 1-proton
wire, the two different regimes can exist in the system depending
on the value of the reorientation energy. In the first {\it
small-$\Omega_R$ regime} (for
$\Omega_R<\Omega_R^{*(2)}=\sqrt{\Delta^2-U \Delta}$), each proton
is located near the outer molecular group and the state
$|8\rangle=|1001\rangle$ has the lowest energy
$\lambda_{3}=\Delta$. As $\Omega_R$ increases and approaches the
"critical" value $\Omega_R^{*(2)}$, the transition to the {\it
large-$\Omega_R$ regime} occurs. In this regime (for
$\Omega_R>\Omega_R^{*(2)}$) the lowest energy levels
$\lambda_{2}=\lambda_{6}=\frac{1}{2}(U-q)$
($q=q_{+}(E=0)=q_{-}(E=0)$) correspond to the states
$|\tilde{2}\rangle$ and $|\tilde{6}\rangle$ in (\ref{U-states}),
with one proton located in the interior of the wire. However, in contrast to
section~\ref{1-prot}, the transition between these two regimes
is $U$-dependent, because $\Omega_R^{*(2)}$ contains the energy of
the D-defect $U$. Fig.~5(a) shows the variation of the proton
occupation numbers $\langle n_{l\nu} \rangle$ with $\Omega_R$ for
$U=9.4$~kcal/mol and $U=1.9$~kcal/mol (plotted in the inset). The
"critical" value $\Omega_R^{*(2)}\approx 3$~kcal/mol for the
repulsion energy $U=9.4$~kcal/mol is larger as compared with
$\Omega_R^{*(2)}\approx 1.5$~kcal/mol for $U=1.9$~kcal/mol. So far
as the repulsion energy $U$ is significant,
$\Omega_R^{*(2)}>\Omega_R^*$. However, as $U \rightarrow 0$,
$\Omega_R^{*(2)}$ approaches the "critical" value $\Omega_R^*$ for
the one-proton case . As we can see from (\ref{U-states}) and
(\ref{sp2}), the ground states $|\tilde{2}\rangle$ and
$|\tilde{6}\rangle$ in the large-$\Omega_R$ regime are represented
by the superpositions of the normal configurations $|7\rangle$ and
$|10\rangle$ and the states $|6\rangle$ and $|11\rangle$
containing the D-defect. Thus the transition to the
large-$\Omega_R$ regime stabilize D-defects inside the chain. The
formation of the D-defect states is clearly observed in Fig.~5(a)
where the occupation numbers of the D-defect states $|6\rangle$
and $|11\rangle$ significantly increase for
$\Omega_R>\Omega_R^{*(2)}$.

Analogously to the one-proton case, the temperature fluctuations
lead to the slight temperature-induced increase of the value
$\Omega_R^{*(2)}(T)$ (corresponding to
$n_{\tilde{3}}=n_{\tilde{2}}=n_{\tilde{6}}$), as compared to
$\Omega_R^{*(2)}$ where the states $|\tilde{2}\rangle$ and
$|\tilde{6}\rangle$ are already stabilized. In fact, for
$\Omega_R>\Omega_R^{*(2)}(T)$ the states corresponding to
protonated chains with significant concentration of D-defects
prevail ($n_{\tilde{2}}=n_{\tilde{6}}>n_{\tilde{3}}$), whereas the
states with the protons located at the boundaries in the
surrounding dominate for $\Omega_R<\Omega_R^{*(2)}(T)$
($n_{\tilde{3}}>n_{\tilde{2}}=n_{\tilde{6}}$). For small $\Omega_R
\ll U$, $\Omega_R^{*(2)}$ can be given by
\begin{equation}\label{omr*2_t}
(\Omega_R^{*(2)}(T))^2=\frac{\Omega_R^{*(2)}}{2}
\left[1+\sqrt{1+\frac{8kTU^2(U/2-\Delta)}{\Delta^2-U\Delta}}
\right]
\end{equation}
Since the line $\Omega_R^{*(2)}(T)$ found from (\ref{omr*2_t}), is
tilted with respect to $T$ in the state diagrams ($T$,
$\Omega_R^{*(2)}(T)$) (Fig.~5(b)), the effect of temperature for
the chain in the large-$\Omega_R$ regime is crucial: with the
increasing $T$, the temperature fluctuations can destroy the D-defects
and redistribute the proton charge between the other chain sites.
See for example the case of $\Omega_R^{*(2)}(T)=3$~kcal/mol and
$U=9.4$~kcal/mol plotted in Fig.~5(b) where the D-defects
annihilate at $T \approx 100$~K.

As $U \rightarrow \infty$, the weight constants for the D-defect
states in (\ref{U-states}) become smaller:
$\frac{\Omega_R}{\sqrt{\Omega_R^2+p_{\pm}^2}}=
\frac{\Omega_R}{\sqrt{\Omega_R^2+\frac{1}{4}(U\pm q )^2}}
\rightarrow 0$. Thus, the contribution of the D-defect-states to
the stable wire configuration goes down as $\frac{1}{U}$ for the
stronger proton repulsion $U$ (see for the comparison $\langle
n_{6} \rangle=\langle n_{11} \rangle$ for different $U$ plotted in
Fig.~5(a)).

The fact that the variation of temperature can lead to formation
or annihilation of the D-defects is also observed in the
$T$-dependence of the proton polarization. Note that especially
for weak external field $E$, the behavior of $P(T)$ in the
small-$\Omega_R$ (Fig.~6(a), $E=0.4 \cdot 10^{7} V/cm$) and in the
large-$\Omega_R$ regime (Fig.~6(b), $E=0.1 \cdot 10^{7} V/cm$) is
drastically different. In the first case, at low $T$, the
predominantly occupied symmetric ground state $|10\ldots 01
\rangle$ has the total polarization $P=0$. However, with $T$
increasing, protons tend to occupy the excited states with
non-symmetric charge distribution that results in an increase of
$P$. Fig.~7(a) demonstrates that the population of all excited
states, in particular those containing D-defects (Fig.~7(a),
inset), grows with $T$. Although the concentration of the D-defect
states $|0011 \rangle$ and $|1100 \rangle$ is of 3-4 orders lower
than that of the normal states (see Fig.~7(a), inset), it
significantly increases up to 1-2 orders with the temperature
increase from $50$ to $300$~K.

In contrast to this, in the large-$\Omega_R$ regime the protons
and stable D-defects migrate in the direction of applied field $E$
for $T \rightarrow 0$ giving a non-zero $P$ (Fig.~6(b)). As $T$
increases, the population of the excited non-defect state with the
protons redistributed at the boundaries grows (see Fig.~7(b))
which gives the lower chain polarization. The D-defects, located
near the end of the chain for finite $E$ ($|0011\rangle$), are
redistributed between another chain positions with $T$, which is
observed in Fig.~7(b) showing a decrease of $|0011 \rangle$-
together with a slight increase of the $|1100 \rangle$-state
population at $T=300$~K as compared to lower temperatures.

We study now the electric field effect in correlated chains.
Fig.~8 shows the variation of polarization and redistribution of
protons with increasing $E$. Consider first the small-$\Omega_R$
regime. In distinct to the 1-proton wire, where the polarization
increases smoothly to its maximal value $P_{max}$ (Fig.~3(a),
inset), we observe here two different threshold effects. The first
transition from the state $|8\rangle$ of (\ref{sp2})(the ground
state of the wire in the small-$\Omega_R$ regime at $E=0$) to the
state $|10\rangle$ (where both of the protons are ordered in the
right position of each H-bond in the direction of the field)
occurs at the threshold field value
\begin{eqnarray}
E_{1}=\frac{-\Delta}{(N-1)(2\mu_r+\mu_{ab})-2\mu_r}.\label{E1}
\end{eqnarray}
The distribution of the occupation probabilities $\langle n_{8}
\rangle$ and $\langle n_{10} \rangle$ for the states $|8\rangle$
and $|10\rangle$ is plotted in Fig.~8(b). We observe at $E=E_1$
the abrupt increase of $\langle n_{10} \rangle$, while at the same
field value $\langle n_{8} \rangle$ drops to zero. Furthermore, we
conclude from (\ref{E1}) that the value $E_1$ lowers with the
number $N$ of the water molecules in the chain. This effect can be
observed in Fig.~9 where the jumps of the polarization are plotted for
different $N$. Finally, for very long water chain ($N \rightarrow
\infty$) $E_1 \rightarrow 0$. In contrast to the strong
$N$-dependence of $E_1$, the second threshold effect appears at
\begin{eqnarray}
E_{2}=\frac{U}{2\mu_r}\label{E2}
\end{eqnarray}
essentially due to the proton correlations and does not depend on
the chain length. The strong increase of $P$ at $E=E_2$ shown in
Fig.~8(a) and Fig.~9 is related to the second drastic
redistribution of the proton charge in the wire. As can be
observed in Fig.~8(b), at $E=E_2$ the occupation probability
$\langle n_{11} \rangle$ of the D-defect-state $|11\rangle$
drastically increases to 1, whereas $\langle n_{10} \rangle$ drops
to zero. Thus, as resulted from our model, the formation of
D-defect in external electric field has a step-like character
proceeding via the threshold mechanism. In the large-$\Omega_R$
regime, where the protons are stabilized at the inner water
molecules already at $E=0$, the first threshold phenomenon at
$E=E_1$, observed for the small-$\Omega_R$ case, does not occur.
However, the transition at $E=E_2$ with the increase of the
D-defect concentration appears in this regime similarly to the
regime of small $\Omega_R$, that can be observed in the
$P$-profile for $\Omega_R=3$~kcal/mol shown in Fig.~8(a).
Note that the effect of the increasing double occupancy due to membrane potentials
has been observed in the current/concentration plots in gramicidin channels \cite{phillips},
thus supporting our main conclusions about the role of the external electric field.

The discussed above formation of the D-defects for
$\Omega_R>\Omega_R^*$ in the high electric field results in the
increase of $P$ for lower temperatures as shown in Fig.~6(b).
Basically, the essential effect of $T$ observed in the
$P(T)$-profiles in Fig.~6, is the suppression of the total
polarization due to proton disorder. However, the shapes of the
polarization in Fig.~6(a) are drastically different for $E<E_1$
and $E>E_2$. For low fields ($E<E_1$) the polarization first
increases (reflecting the fluctuation-induced expansion of proton
charge from the outer symmetric positions $|10 \ldots 01 \rangle$
with $P=0$ to the inner positions of the chain accompanied by the
formation/annihilation of D-defects), and then smoothly decreases
due to the disorder effect. In contrast to this, as the increasing
electric field induces the step-like formation of D-defects in the
small-$\Omega_R$ regime, the temperature behavior of $P$ in this
case is similar to the the large-$\Omega_R$ case (compare
Fig.~6(a) and Fig.~6(b) with $E=1\cdot 10^{7}$~V/cm) showing the
smooth disorder-induced decrease of $P$ with $T$.

We also note that the stable configurations with double proton occupancy require the additional
reorientation steps for the proton translocation and can result in the smaller values
for the proton conductivity. This fact has been observed in the measurements of the proton
conductance in two different stereoisomers of the gramicidin \cite{cukierman}, thus supporting 
a possibility of stabilization of the D-defect states in proton wires.

\section{SUMMARY}

In this work we studied the process of proton translocation in
1D-chains mimicking protonated water channels embedded in
surrounding. We have analyzed the role of the reorientation motion
of protons, as well as the effect of electric field and proton
correlations on the chain dynamics. We have shown that the
increase of the reorientation energy results in the transition to
the large-$\Omega_R$ regime characterized by the transfer of the
proton charge from the surrounding to the inner water molecules in
the chain. The process of proton migration along the chain in the
external electric field has the step-like character leading to the
appearance of the electric field threshold-type phenomena with
drastic redistribution of proton charge. The correlations between
protons in the chain increase the "critical" reorientation energy
$\Omega_R^*$ necessary for the transition into the
large-$\Omega_R$ regime, where the protonated chain contains a
finite concentration of Bjerrum defects. The temperature
fluctuations induce a slight increase of $\Omega_R^*(T)$
separating the state with the protons located in surrounding near
the outer groups, and the protonated state with D-defects. For the
correlated chains, this temperature dependence of the "critical"
reorientation energy can lead to the redistribution of proton
charge and annihilation of D-defects with increasing $T$. The
electric field applied to the correlated chains induces first the
formation of ordered dipole structures for the lower $E$ values,
and than, with the further $E$ increase, the stabilization of
the states with the Bjerrum D-defects.

Generally, the increase of temperature suppresses the total
polarization in the chain due to the increasing disorder. However,
especially in the low electric fields, the shapes of the
temperature profiles of the polarization appear to be drastically
different in the small- and large-$\Omega_R$ regimes demonstrating
the complex interplay between the reorientation energy and
temperature.

Finally, as follows from our analysis, the following factors
strongly influence the formation of Bjerrum defects: (i) the high
electric fields can form the defects and pump them in the chain in
the direction of field; (ii) the increase of the orientational energy barrier
leads to the stabilization of D-defects; (iii) the increase of
temperature in the large-$\Omega_R$ regime results in the
formation/annihilation of D-defects, whereas for small $\Omega_R$
the concentration of D-defects significantly increases up to 1-2
orders at the room temperatures as compared to the low $T \approx
50$~K.

\newpage

\appendix
\section{DECOMPOSITION OF THE PROTON STATES IN THE SYSTEM WITH $N=2$ H-BONDS}

We demonstrate below the procedure of the mapping in the system
with $N=2$ H-bonds on the multi-site states. For $N=2$ the basis
$|i \rangle$ includes $2^{2N}=16$ states
$|n_{1a},n_{1b},n_{2a},n_{2b} \rangle$:
\begin{eqnarray}
|1\rangle=|0000\rangle, & |2\rangle=|1000\rangle, & |3\rangle=|0100\rangle,\nonumber\\
|4\rangle=|0010\rangle, & |5\rangle=|0001\rangle, & |6\rangle=|1100\rangle, \label{sp2}\\
|7\rangle=|1010\rangle, & |8\rangle=|1001\rangle, & |9\rangle=|0110\rangle,\nonumber \\
|10\rangle=|0101\rangle, & |11\rangle=|0011\rangle, & \ldots
|16\rangle=|1111\rangle.\nonumber
\end{eqnarray}

We can derive the relations between $c_{l,\nu}$ and $X^{ii'}=|i
\rangle \langle i'|$:
\begin{eqnarray}
c_{l,\nu}=\sum_{i,j} \langle i|c_{l,\nu}|j \rangle X^{ij},
\label{ferm_hubb}
\end{eqnarray}
where the expectation numbers $\langle i|c_{l,\nu}|j \rangle$ can
be found using the usual antisymmetric rules for Fermi-operators
\cite{davydov}. Specifically, for the case $N=2$ the expressions
(\ref{ferm_hubb}) yield:
\begin{eqnarray}
&&c_{0,a}=X^{1,2}+X^{3,6}+X^{4,7}+X^{5,8}+X^{9,12}+X^{10,13}+X^{11,14}+X^{15,16}\nonumber\\
&&c_{1,a}=X^{1,4}-X^{2,7}-X^{3,9}+X^{5,11}+X^{6,12}-X^{8,14}-X^{10,15}+X^{13,16}\label{c22}\\
&&c_{0,b}=X^{1,3}-X^{2,6}+X^{4,9}+X^{5,10}-X^{7,12}-X^{8,13}+X^{11,15}-X^{14,16}\nonumber\\
&&c_{1,b}=X^{1,5}-X^{2,8}-X^{3,10}-X^{4,11}+X^{6,13}+X^{7,14}+X^{9,15}-X^{12,16}\nonumber
\end{eqnarray}
Using the relations (\ref{c22}) and the fact that $X^{ii'}
X^{ll'}=\delta_{i'l}X^{il'}$ (due to the orthogonality of the
states $|i \rangle$), we decompose (\ref{h_t}-\ref{h_e}) in terms
of $X^{ii'}$ operators into the following 5 terms:
\begin{eqnarray}
{H}={H_{2}^0}\oplus{H_{2}^1}\oplus{H_{2}^2}\oplus
{H_{2}^3}\oplus{H_{2}^4}, \label{ham_decomp}
\end{eqnarray}
where \noindent
\begin{eqnarray}
{H_{2}^{0}}&=&a_{2}^{0},\label{ham21}\nonumber\\
{H_{2}^{1}}&=&(\Delta+(\mu_r+\mu_{ab})E)X^{2,2}+\mu_r E(X^{3,3}-X^{4,4})+\nonumber \\
&&(\Delta-(\mu_r+\mu_{ab})E)X^{5,5}+\Omega_T(X^{2,3}+X^{3,2})+ \\
&&\Omega_T(X^{4,5}+X^{5,4})+\Omega_R(X^{3,4}+X^{4,3})+
a_{2}^{1},\nonumber
\end{eqnarray}
\begin{eqnarray}
{H_{2}^{2}}&=&(U+(2\mu_r+\mu_{ab})E)X^{6,6}+\mu_{ab}(X^{7,7}-X^{10,10})+ \label{ham22}\nonumber \\
&&\Delta X^{8,8}+(J-\Delta)X^{9,9}+(U-(2\mu_r+\mu_{ab})E)X^{11,11}+\nonumber \\
&&\Omega_T(X^{7,8}+X^{8,7})+\Omega_T(X^{8,10}+X^{10,8})+\Omega_T(X^{9,10}+X^{10,9})+ \nonumber \\
&&\Omega_R(X^{6,7}+X^{7,6})+\Omega_R(X^{10,11}+X^{11,10})+a_{2}^{2},
\end{eqnarray}
\begin{eqnarray}
{H_{2}^{3}}&=&(U+J-\Delta+(\mu_r+\mu_{ab})E)X^{12,12}+(U+\mu_r E)X^{13,13}+ \label{ham23}\nonumber \\
&&(U-\mu_r E)X^{14,14}+(U+J-\Delta-(\mu_r+\mu_{ab})E)X^{15,15} +\\
&&\Omega_T(X^{12,13}+X^{13,12})+\Omega_T(X^{14,15}+X^{15,14})+\nonumber \\
&&\Omega_R(X^{13,14}+X^{14,13})+a_{2}^{3},\nonumber \\
{H_{2}^{4}}&=&(2U+J)X^{16,16}+a_{2}^{4}\nonumber.
\end{eqnarray}

Since the parameter
\begin{equation} \label{delta}
\Delta=(\tilde{w}-\tilde{\varepsilon})-(\varepsilon-w')
\end{equation}
in (\ref{ham21})-(\ref{ham23}) is the difference between proton
configuration energies at the boundary ($l=1$ or $l=N+1$
surrounding molecular groups), and at the inner ($2<l<N$) water
molecule, it describes, in fact, the energy barrier for the
protonation of the water chain. For our analysis $\Delta$ has the
key importance, because the other energy constants in
(\ref{ham21})-(\ref{ham23})
\begin{eqnarray*}
a_2^0=a_2^2-(\tilde{w}-\tilde{\varepsilon})-(\varepsilon-w'), &&
a_2^1=a_2^2-(\tilde{w}-\tilde{\varepsilon}),\nonumber \\
a_2^3=a_2^2+(\tilde{w}-\tilde{\varepsilon}), &&
a_2^4=a_2^2+(\tilde{w}-\tilde{\varepsilon})+(\varepsilon-w'),\nonumber\\
a_2^2=\tilde{w}+\tilde{\varepsilon}+\varepsilon && \nonumber
\end{eqnarray*}
which appear due to the boundary effects, are independent of the
proton location in the wire and thus do not influence the
statistical characteristics like (\ref{pp}).

By the similar way the energy (\ref{h_t}-\ref{h_e}) can be
rewritten for the systems with any finite value of $N$.






\begin{figure}[htbp]
\epsfxsize=8.0cm \centerline{\epsffile{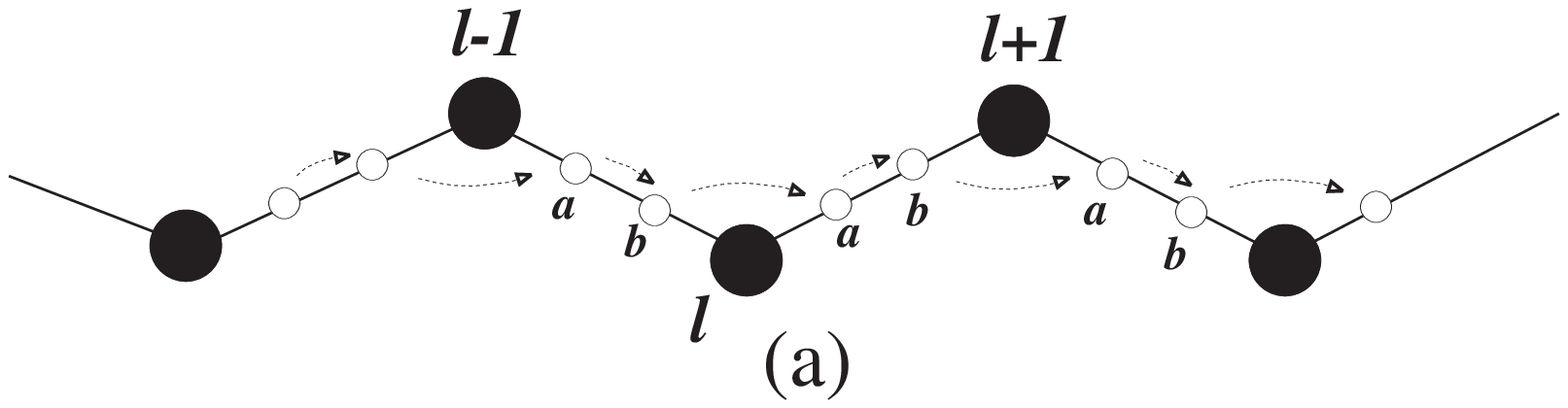}}
\null\vspace{0.2in} \epsfxsize=8.0cm
\centerline{\epsffile{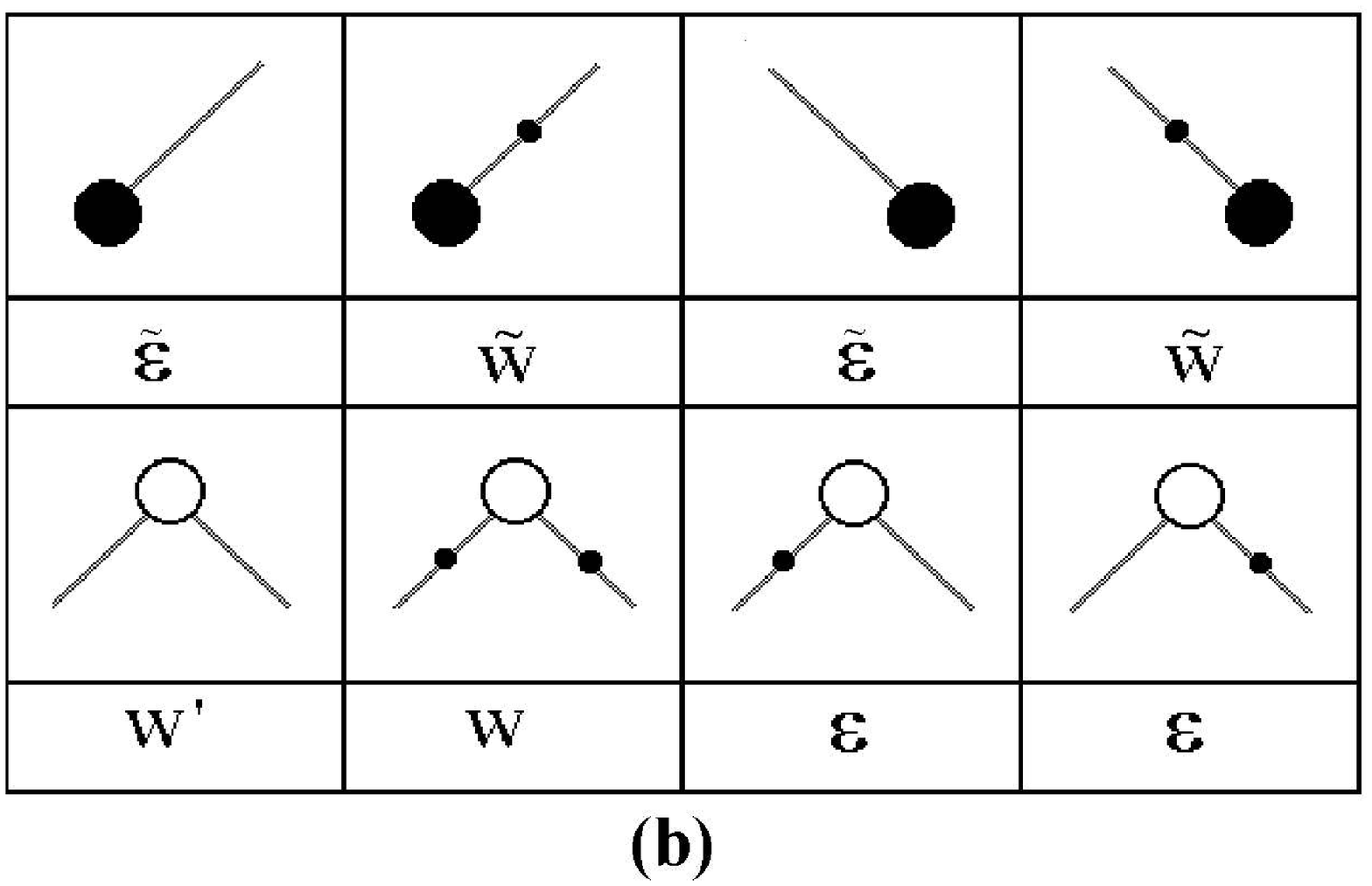}} \null\vspace{4.in}
\null\vspace{0.2in}

\caption{\label{fig1} (a) Schematic presentation of proton wire,
arrows indicate a possible path of proton migration along the
chain. Full circles denote water molecules and open circles are
the possible positions for excess proton. (b) Scheme of possible
proton configurations near the outer surrounding groups (the upper
part) and the inner water molecules of the wire (the lower part).}
\end{figure}

\begin{figure}[htbp]
\epsfxsize=8.0cm \centerline{\epsffile{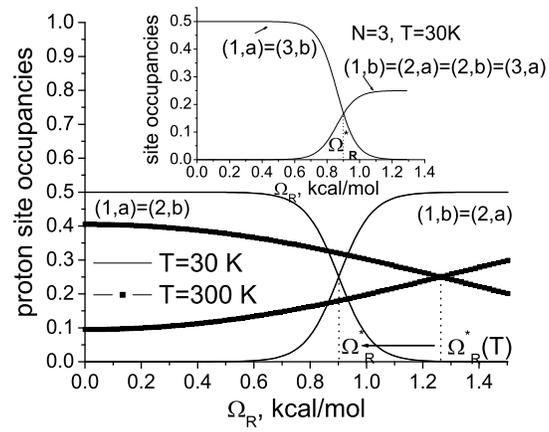}}
\null\vspace{0.2in}

\caption{\label{fig2}Proton site occupancies for different
reorientation energies $\Omega_R$ in the chain with two hydrogen
bonds containing one excess proton. The inset shows the
redistribution of proton charge in the chain with $N=3$ H-bonds.}
\end{figure}

\begin{figure}[htbp]
\epsfxsize=8.0cm \centerline{\epsffile{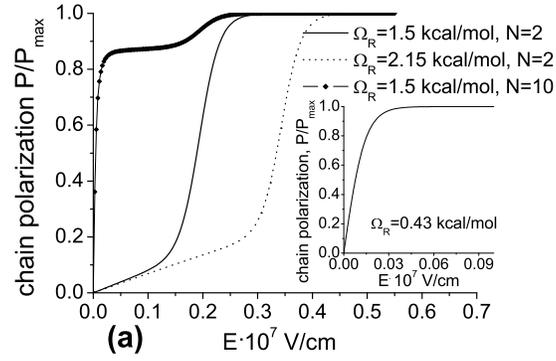}}
\null\vspace{0.2in} \epsfxsize=8.0cm
\centerline{\epsffile{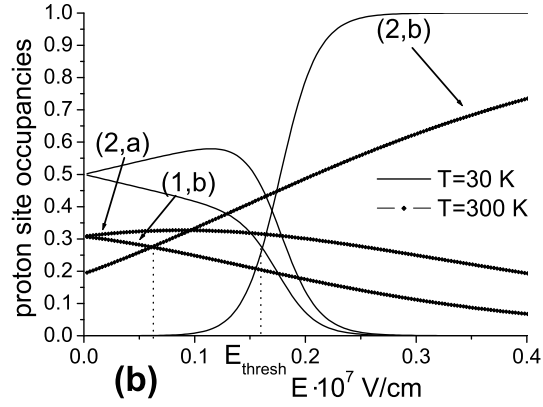}}

\null\vspace{0.2in}

\caption{\label{fig3} (a) Proton polarization vs electric field in
the small-$\Omega_R$ (shown in the inset)- and large-$\Omega_R$
regimes for $T=30 K$, and (b) average site occupancies vs $E$ in
the chain containing $N=2$ hydrogen bonds and $n=1$ proton for
$\Omega_R=1.5$~kcal/mol and for different temperatures.}
\end{figure}

\begin{figure}[htbp]
\epsfxsize=8.0cm \centerline{\epsffile{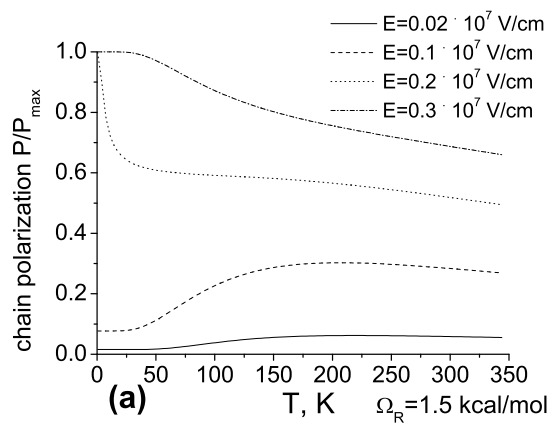}}
\null\vspace{0.2in} \epsfxsize=8.0cm
\centerline{\epsffile{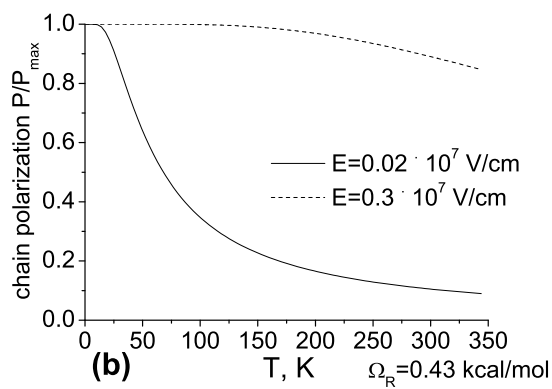}}

\null\vspace{0.2in}

\caption{\label{fig4}Proton polarization vs temperature (a) in the
large-$\Omega_R$ regime and (b) in the small-$\Omega_R$ regimes
for different values of applied electric field $E$.}
\end{figure}

\begin{figure}[htbp]
\epsfxsize=8.0cm \centerline{\epsffile{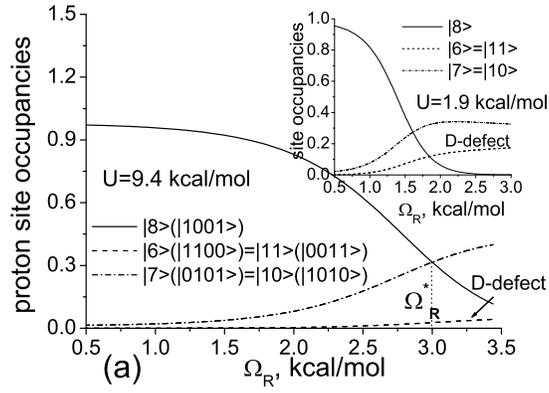}}
\null\vspace{0.2in} \epsfxsize=8.0cm
\centerline{\epsffile{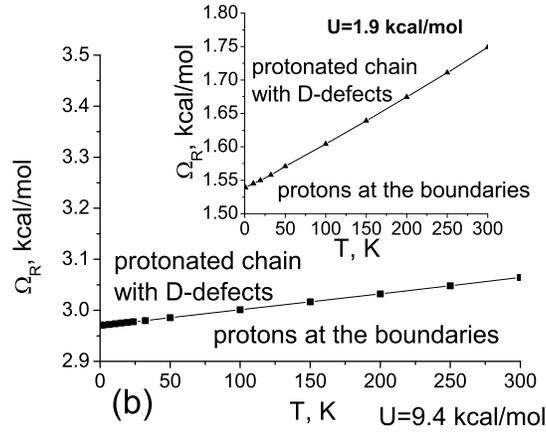}}

\null\vspace{0.2in}

\caption{\label{fig5}(a) Proton charge distribution vs $\Omega_R$
in the H-bonded chain ($N=2$) containing two protons ($n=2$) for
$T=100K$. The inset shows the variation of proton charge with
$\Omega_R$ for $U=1.9$~kcal/mol. (b) State diagrams ($T$,
$\Omega_R$) for $N=2$, $n=2$ indicating the regions of stability
of the protonated chain with D-defect and the states with the
protons localized at the boundaries. The inset shows the state
diagram for lower $U=1.9$~kcal/mol.}
\end{figure}

\begin{figure}[htbp]
\epsfxsize=8.0cm \centerline{\epsffile{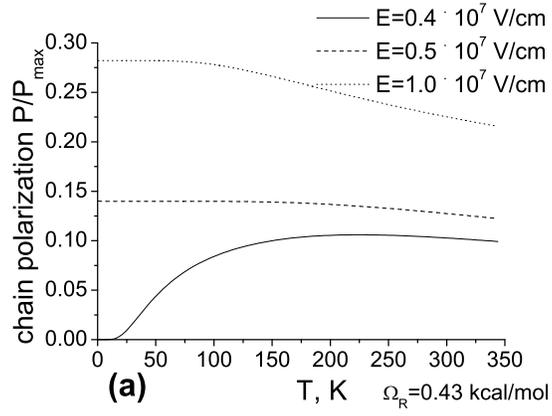}}
\null\vspace{0.2in} \epsfxsize=8.0cm
\centerline{\epsffile{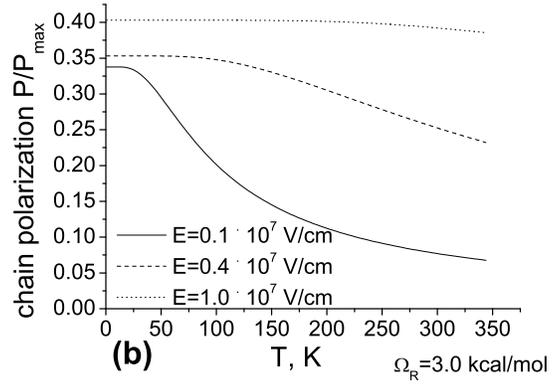}}

\null\vspace{0.2in}

\caption{\label{fig6}Proton polarization in the chain with $N=2$
H-bonds and $n=2$ protons vs temperature (a) in the
small-$\Omega_R$ regime and (b) in the large-$\Omega_R$ regimes
for different values of applied electric field $E$.}
\end{figure}

\begin{figure}[htbp]
\epsfxsize=8.0cm \centerline{\epsffile{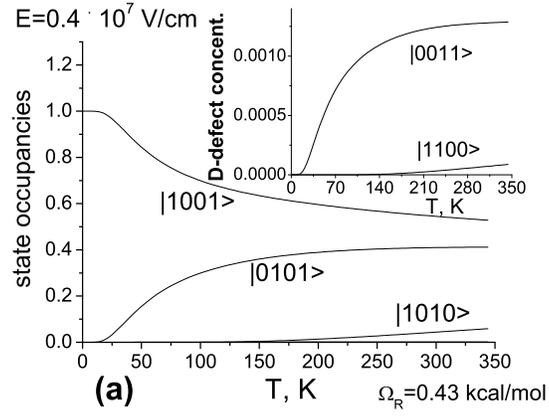}}
\null\vspace{0.2in} \epsfxsize=8.0cm
\centerline{\epsffile{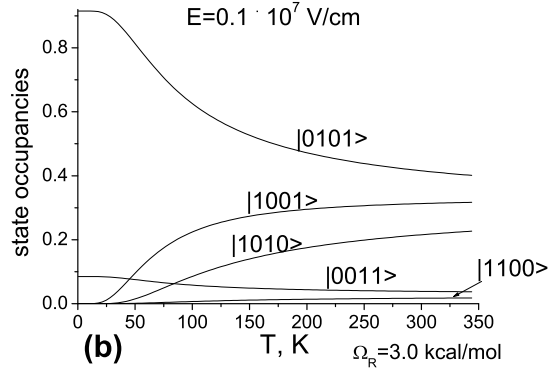}}

\null\vspace{0.2in}

\caption{\label{fig7}Proton site occupancies in the chain
containing $N=2$ H-bonds and $n=2$ protons vs temperature (a) in
the small-$\Omega_R$ regime and (b) large-$\Omega_R$ regimes for
low electric field $E$. The inset shows the concentration of
D-defects vs $T$ in the small-$\Omega_R$ regime.}
\end{figure}

\begin{figure}[htbp]
\epsfxsize=8.0cm \centerline{\epsffile{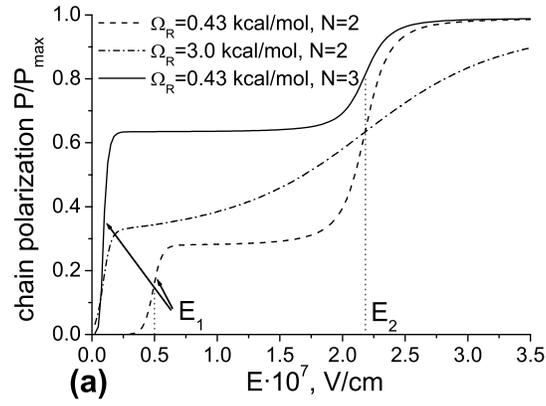}}
\null\vspace{0.2in} \epsfxsize=8.0cm
\centerline{\epsffile{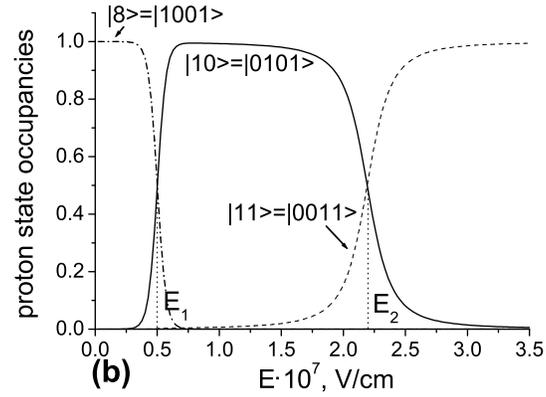}}

\null\vspace{0.2in}

\caption{\label{fig8}(a) Proton polarization for different
$\Omega_R$ and (b) average site occupancies for
$\Omega_R=0.43$~kcal/mol vs electric field $E$ in the chain
containing $N=2$ hydrogen bonds and $n=2$ protons at $T=30K$.}
\end{figure}


\begin{figure}[htbp]
\epsfxsize=8.0cm \centerline{\epsffile{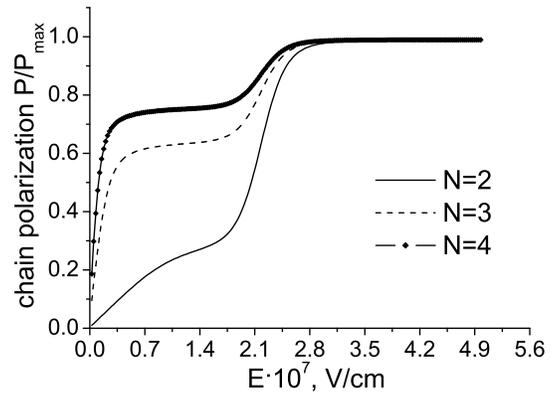}}

\null\vspace{0.2in}

\caption{\label{fig9}Proton polarization vs $E$ in the chains of
different length containing two protons for
$\Omega_R=0.43$~kcal/mol and $T=300K$.}
\end{figure}

\end{document}